\PassOptionsToPackage{dvipsnames,table}{xcolor}
\documentclass[runningheads,anonymous]{llncs}

\usepackage[T1]{fontenc}
\usepackage{graphicx}

\usepackage{makecell}
\usepackage{multirow}
\usepackage{longtable}
\usepackage{float}
\usepackage{tikz}
\usepackage{amssymb}
\usepackage{booktabs}

\usepackage{soul}

\usepackage{bm}
\usepackage{amsmath}

\usepackage[hidelinks]{hyperref}
\usepackage{url}

\usepackage{cleveref}
\usepackage[nocompress]{cite}

\usepackage{comment}

\usepackage{listings}

\usepackage{pifont}
\newcommand{\xmark}{\ding{55}}%

\makeatletter
\let\oldhypertarget\hypertarget
\renewcommand{\hypertarget}[2]{%
  \oldhypertarget{#1}{#2}%
    \protected@write\@mainaux{}{%
        \string\expandafter\string\gdef
          \string\csname\string\detokenize{#1}\string\endcsname{#2}%
    }%
  }
\newcommand{\myhyperlink}[1]{%
  \hyperlink{#1}{\csname #1\endcsname}%
  }
\makeatother

\newcommand\encircle[1]{%
  \tikz[baseline=(X.base)] 
    \node (X) [draw, shape=circle, inner sep=0] {\strut #1};}

\newcommand{\midsepremove}{\aboverulesep = 0mm \belowrulesep = 0mm}
\midsepremove
\newcommand{\midsepdefault}{\aboverulesep = 0mm \belowrulesep = 0mm}
\midsepdefault

\newenvironment{rulefont}{\fontfamily{pcr}\selectfont}{\par}
\DeclareTextFontCommand{\textrulefont}{\rulefont}

\newcolumntype{L}[1]{>{\raggedright\let\newline\\\arraybackslash\hspace{0pt}}m{#1}}
\newcolumntype{C}[1]{>{\centering\let\newline\\\arraybackslash\hspace{0pt}}m{#1}}
\newcolumntype{R}[1]{>{\raggedleft\let\newline\\\arraybackslash\hspace{0pt}}m{#1}}

\newcommand{\mynote}[3]{
\fbox{\bfseries\sffamily\scriptsize#1}
{\small\textsf{\emph{\color{#3}{#2}}}}}

\newcommand{\johan}[1]{\mynote{Johan}{#1}{teal}}
\newcommand{\lucas}[1]{\mynote{Lucas}{#1}{red}}
\newcommand{\gilles}[1]{\mynote{Gilles}{#1}{blue}}
\newcommand{\pierre}[1]{\mynote{Pierre}{#1}{orange}}

\renewcommand{\johan}[1]{}
\renewcommand{\lucas}[1]{}
\renewcommand{\gilles}[1]{}
\renewcommand{\pierre}[1]{}

\begin{document}

\title{Overlapping$\:$data$\:$in$\:$network$\:$protocols:$\:$bridging OS$\:$and$\:$NIDS$\:$reassembly$\:$gap}

\author{Lucas Aubard\inst{1} \and
    Johan Mazel\inst{2} \and
    Gilles Guette\inst{3} \and
    Pierre Chifflier\inst{2}
}
\institute{}
\institute{
    Inria, Rennes, France \email{lucas.aubard@inria.fr} \and
    ANSSI, Paris,France \email{firstname.lastname@ssi.gouv.fr}\and
    IMT Atlantique, Cesson Sévigné, France \email{gilles.guette@imt-atlantique.fr}
}

\maketitle              

\begin{abstract}
    IPv4, IPv6, and TCP have a common mechanism allowing one to split an original data packet into several chunks.
    Such chunked packets may have overlapping data portions and, OS network stack implementations may reassemble these overlaps differently.
    A Network Intrusion Detection System (NIDS) that tries to reassemble a given flow data has to use the same reassembly policy as the monitored host OS;
    otherwise, the NIDS or the host may be subject to attack.
    In this paper, we provide several contributions that enable us to analyze NIDS resistance to overlapping data chunks-based attacks.
    First, we extend state-of-the-art \emph{insertion} and \emph{evasion} attack characterizations to address their limitations in an overlap-based context.
    Second, we propose a new way to model overlap types using Allen's interval algebra, a spatio-temporal reasoning.
    This new modeling allows us to formalize overlap test cases, which ensures exhaustiveness in overlap coverage and eases the reasoning about and use of reassembly policies.
    Third, we analyze the reassembly behavior of several OSes and NIDSes when processing the modeled overlap test cases.
    We show that
    1) OS reassembly policies evolve over time and
    2) all the tested NIDSes are (still) vulnerable to overlap-based evasion and insertion attacks.
    \keywords{Intrusion detection system \and IP \and TCP \and Overlapping data}
\end{abstract}



\section{Introduction}
\label{sec:introduction}

Some Internet protocols use chunking\footnote{We use the term chunking as a generic way to refer to "splitting an original data chunk into several". Thus, it both refers to the "fragmentation" mechanism for IPv4 and IPv6 and the "segmentation" mechanism for TCP.} mechanism.
It was introduced to answer a potential discrepancy between media link capacity at a node's entry and a node's exit along the path between a sender and a receiver.
When chunking occurs, the receiver must reassemble all the chunks to retrieve the initial data packet.
However, the chunking mechanism can lead to overlaps.
The most common case is a chunk retransmission with the same data that starts and finishes at the same byte offsets.
Nevertheless, other types of overlaps exist, i.e., partial overlaps, and the data can be different on the overlapping portion. IPv4 and TCP RFC specifications~\cite{rfc791,rfc9293} neither forbid data overlaps nor specify the behavior an implementation must adopt (e.g., prefer data from the older chunk).
IPv6 RFC specification initially did not forbid overlaps in the first drafts,
but since 2017, it has banned them~\cite{deering2017rfc}.

Network packet analysis is one of the possible techniques commonly used to detect intrusions.
Some widely deployed Network Intrusion Detection Systems (NIDS) are signature-based, meaning that they match suspicious patterns or signatures of known attacks on the reassembled flow data.
Therefore, NIDSes must reassemble consistently the network traffic with the monitored hosts to detect attacks. 
Ptacek and Newsham~\cite{ptacek1998insertion} introduced a set of IP and TCP ambiguities that may lead to NIDS misassemblies with supervised hosts and thus, NIDS circumvention.
The ambiguities exist because
1) NIDSes receive a copy of the network traffic from and to the hosts, and
2) NIDSes and monitored hosts are distinct machines.
Thus, NIDSes cannot easily determine how a host processes a specific packet when data overlap occurs.

\begin{figure}[!htb]
    \centering
    \includegraphics[width=0.8\columnwidth]{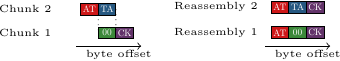}
    \caption{\label{fig:overlap} Data overlap ambiguity illustration.}
\end{figure}

\Cref{fig:overlap} is an illustration of the data overlap issue.
The reassembled data here differs depending on which chunk the reassembly policy favors.
Someone with bad intentions may exploit the multiple reassembly possibilities to hide a malicious payload.
If the NIDS reassembles with reassembly strategy 1 while the host reassembles with strategy 2, the former cannot see the malicious payload and raise any security alert.

Several works showed that reassemblies depend on the IPv4~\cite{khattak2013towards,novak2005target,ptacek1998insertion,shankar2003active}, IPv6~\cite{atlasis2012attacking,di2023new,lin2024research} and TCP~\cite{khattak2013towards,novak2007target,ptacek1998insertion,shankar2003active} implementations.
Since attackers may use the overlapping ambiguity to bypass their security functionalities, well-known NIDSes like Suricata~\cite{suricata} 
and Snort~\cite{roesch1999snort} introduced a feature allowing users to associate each supervised host (through IP address) to a specific reassembly policy~\cite{rpsuricata,rpipsnort,rptcpsnort}.
Other NIDSes like Zeek (formerly Bro~\cite{paxson1999bro}) have chosen a different approach: implementing only one reassembly policy but allowing users to enable overlap-related alerts.
The set of implemented IPv4 and TCP reassembly policies in Suricata and Snort are based on the works of Novak and Sturges~\cite{novak2005target,novak2007target} published in 2005 and 2007.
These works are the latest that tested OS policies for IPv4 and TCP.
Thus, we identify two main problems.
The first one is that the OSes Novak and Sturges tested have since been updated, and their protocol implementation may have changed.
The second problem arises from the manual approach the related works~\cite{ptacek1998insertion,shankar2003active,novak2005target,novak2007target,atlasis2012attacking,di2023new} used to design their test cases.
There is thus no certainty on test case coverage exhaustiveness.
Reassembly policies implemented within Suricata and Snort may be based on out-of-date and/or partial policy descriptions, giving a wrong sentiment of security regarding overlap-based attacks.
So, the knowledge of modern OS reassembly policies must be updated (see~\Cref{sec:os_reassemblies}), and the exhaustive coverage of overlap test cases must be ensured (see~\Cref{sec:test_case_modeling,sec:related_works}).

After Ptacek and Newsham's seminal work~\cite{ptacek1998insertion}, an entire research area has focused on finding any sequence of packets (i.e., not only based on the overlap ambiguity) that protocol implementations process differently.
Research has especially intensified from the growing deployment of censorship systems (CS) since the technique has been found successful in their circumvention~\cite{bock2019geneva,bock2021even,khattak2013towards,li2017lib,wang2017your,wang2020symtcp}.
The methods used to find such sequences have moved from manual~\cite{khattak2013towards,li2017lib,wang2017your} (in which authors have to discover the ambiguities all by themselves) to semi-automatic ones using fuzzing~\cite{bock2019geneva,zhang2022statediver,zou2021tcp} or symbolic execution~\cite{wang2020symtcp,wang2021themis}.
Until 2020, some works~\cite{khattak2013towards,li2017lib,wang2017your,wang2020symtcp} reported the (more and more) relative success of overlapping chunk-based attacks.
We argue that the recent works using semi-automatic approaches did not perform extensive (i.e., complete) testing using overlap-based strategies, mainly because fuzzing and symbolic execution methods are good at finding novel chunk sequence examples but not at exhaustive testing.

To our knowledge, no work has verified that NIDSes' overlap reassembly policies are consistent with OSes' since Ptacek and Newsham unveiled the issue in 1998.
We fill that gap in this paper.
The question that will guide us throughout is: \emph{Do NIDSes reassemble overlap test cases differently as the OSes, and therefore, is it possible to use data overlaps to attack a NIDS or the hosts it supervises?}
The contributions are the following:
\begin{itemize}
    \item In~\Cref{sec:threat_model_problem_defintion}, we extend \emph{insertion} and \emph{evasion} definitions to overlap-based attack context.
          This enables us to cover all the related attack scenarii and to characterize the requirements for the overlapping data portion.
    \item In~\Cref{sec:chunk_sequence_modeling}, we model chunk sequences with Allen's spatio-temporal reasoning~\cite{allen1983maintaining} and, thus, ensure overlap coverage exhaustiveness (in contrast to 10 over 13 related-works, see~\Cref{sec:related_works}).
    \item In~\Cref{sec:results}, we describe IPv4, IPv6, and TCP reassembly policies of a large range of OSes, including recent ones (e.g., Windows 11, the Linux-based Debian 12, FreeBSD 14.1, OpenBSD 7.6, Solaris 11.4) and of three widely deployed NIDSes (i.e., Snort, Suricata and Zeek). 
          We find that:  
          \begin{itemize}
              \item [i)] OS reassembly policies evolve over time.
                    In particular, Windows and Linux-based OSes have changed their IPv4 reassembly policies (regarding state-of-the-art), while TCP policies have barely been modified.
              \item [ii)] Snort, Suricata, and Zeek reassemble IPv4, IPv6, and TCP chunks in a partially consistent way with OSes.
                    This opens the way to insertion and evasion attacks.
                    A CVE~\cite{cve_ours} was assigned to some of the disclosed problems.
          \end{itemize}
\end{itemize}

\section{Problem Definition and Threat Model}
\label{sec:threat_model_problem_defintion}

This section first defines insertions and evasions in the specific context of overlap-based attacks.
It then details the attacker capabilities needed to exploit OS and NIDS reassembly discrepancies.

\subsection{Problem Definition}
\label{sec:problem_defintion}

An attacker may use overlap data ambiguity to exploit NIDS and host reassembly divergence.
We extend Ptacek and Newsham~\cite{ptacek1998insertion} and Wang et al.~\cite{wang2020symtcp} \emph{insertion} and \emph{evasion} packet-based attack definitions to fit the context of IP and TCP chunk overlaps.
The main shortcomings that we address are:
\begin{itemize}
    \item Ptacek and Newsham consider a data chunk as being either \emph{totally} accepted or dropped but not \emph{partially} accepted.
          With the overlap from~\Cref{fig:overlap}, the "ATTACK"/"AT00CK" reassembly divergence is impossible.
    \item Wang et al. do not consider on purpose malicious (or "filtered") payload insertion inside the NIDS flow data, nor do they consider IP-based attacks.
\end{itemize}
We treat the NIDSes and the host OSes as black boxes in the following.
We consider the IP and TCP data stream pushed to the upper layer; thus, our definitions apply to IP and TCP overlap-based attacks.

Let $P = \{$IPv4,$\:$IPv6,$\:$TCP$\}$ be the Internet protocol set with a chunking mechanism we target.
Let $C^p$ be the set of all possible chunks for protocol $p \in P$.

\begin{definition}[$p$ protocol data buffer synchronization]
    \label{def:state_machine_synchro}
    Given a chunk sequence $c_{f}...c_{l} \in C^p$, with $c_f$ (resp. $c_l$) the first (resp. last) sent chunk, we say that the NIDS and the supervised host have their $p$ data buffers synchronized if the next upper-layer data streams are the same.
\end{definition}

The $p$'s data buffer desynchronization requires
chunks linked with a particular overlap type that the NIDS and the host reassemble differently (see~\Cref{sec:results})
and
carefully crafted overlapping data (see~\Cref{sec:threat_model}). 
Such data is either an \emph{evasion} or an \emph{insertion} payload depending on the target (i.e., the host or the NIDS).
\emph{Evasion and insertion attacks aim to desynchronize NIDS and host $p$ protocol data buffers}.

\begin{definition}[Evasion in a data overlap-based context]
    \label{def:evasion}
    An evasion attack consists of \emph{some malicious payload} that is not visible in the data analyzed by the NIDS while it is visible on the \emph{supervised host}'s reassembled payload.
\end{definition}

\begin{definition}[Insertion in a data overlap-based context]
    \label{def:insertion}
    Symmetrically, an insertion attack consists of \emph{some malicious payload} that is visible in the flow data analyzed by the \emph{NIDS} while it is not on the host's reassembled payload.
\end{definition}

\subsubsection{Illustration}
\label{sec:attack_type_illustration}

\renewcommand{\arraystretch}{1}
\begin{table}[!t]
    \setlength\tabcolsep{3pt}
    \centering
    \begin{tabular}{l l c c c c c c}
        \toprule
        \multirow{2}{*}{\textbf{Attack type}} & \multirow{2}{*}{\textbf{Implem.}} & \multirow{2}{*}{\textbf{Target}} & \multirow{2}{*}{\textbf{\makecell{Reassembled                                                                                                                                                \\data}}} &   \multirow{2}{*}{\textbf{\makecell{Attack\\scenario}}} & \multicolumn{3}{c}{\textbf{\makecell{Works}}} \\
        \cline{6-8}
                                              &                                   &                                  &                                               &                                       & \textbf{\cite{ptacek1998insertion}} & \textbf{\cite{wang2020symtcp}} & \textbf{Us}                   \\
        \midrule

        \multirow{4}{*}{Evasion}              & NIDS                               &                                  & -                                             & \multirow{2}{*}{\hypertarget{E1}{E1}} & \multirow{2}{*}{$\checkmark$}       & \multirow{2}{*}{$\checkmark$}  & \multirow{2}{*}{$\checkmark$} \\
                                              & Sup. host                         & \xmark                           & "ATTACK"                                      &                                                                                                                                              \\
        \cline{2-8}
                                              & NIDS                               &                                  & "AT00CK"                                      & \multirow{2}{*}{\hypertarget{E2}{E2}} &                                     & \multirow{2}{*}{$\checkmark$}  & \multirow{2}{*}{$\checkmark$} \\
                                              & Sup. host                         & \xmark                           & "ATTACK"                                      &                                                                                                                                              \\
        \hline
        \multirow{4}{*}{Insertion}            & NIDS                               & \xmark                           & "ATTACK"                                      & \multirow{2}{*}{\hypertarget{I1}{I1}} & \multirow{2}{*}{$\checkmark$}       &                                & \multirow{2}{*}{$\checkmark$} \\
                                              & Sup. host                         &                                  & -                                             &                                                                                                                                              \\
        \cline{2-8}
                                              & NIDS                               & \xmark                           & "ATTACK"                                      & \multirow{2}{*}{\hypertarget{I2}{I2}} &                                     &                                & \multirow{2}{*}{$\checkmark$} \\
                                              & Sup. host                         &                                  & "AT00CK"                                      &                                                                                                                                              \\
        \bottomrule
    \end{tabular}
    \caption{\label{tables/reassembly_cases_numbered} Attack types illustrated with~\Cref{fig:overlap} reassembly cases and based on~\Cref{def:insertion,def:evasion}.
        - means the implementation \emph{ignores} the flow chunk data.
    }
\end{table}

We use the overlap chunk sequence introduced in~\Cref{fig:overlap} to illustrate insertion and evasion attack types in~\Cref{tables/reassembly_cases_numbered}.
As we can see, the non-targeted host either reassembles differently (with a benign payload, for example) or completely ignores the chunk sequence.
Data buffer desynchronization is one of the most dreaded risks for a NIDS since it eventually allows an attacker to bypass all its security mechanisms.
See related CVE 2019-18625, 2019-18792, or 2021-37592, whose scores are high or critical.
On the one hand, attackers can use evasions to circumvent the NIDS inspection function.
An alert pattern can thus reach the supervised host without the NIDS noticing it, as~\cite{ptacek1998insertion,wang2020symtcp} did.
An evasion can also impact other NIDS functionalities, such as file or TLS certificate extractions, since a unique (overlapped) bit is sufficient to corrupt them.
On the other hand, the insertion attack can alter the NIDS's normal behavior, for instance, by exploiting a known NIDS vulnerability
(e.g., 2019-12175, 2023-7242, or 2024-47522 CVEs).
Or it can raise false positive alerts, wasting analysts' time.

\subsection{Threat Model}
\label{sec:threat_model}

\begin{figure}[!t]
    \centering
    \includegraphics[width=0.8\textwidth]{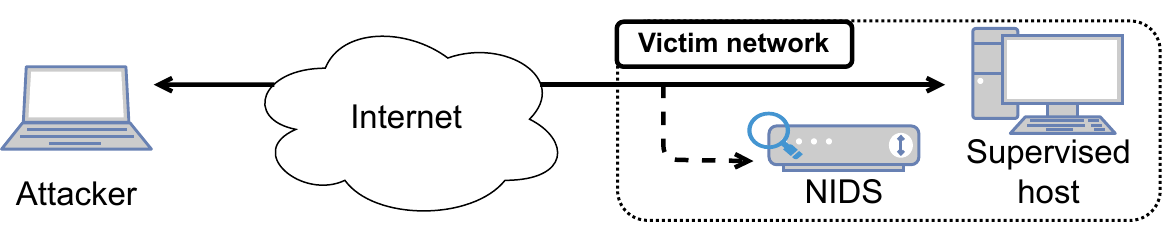}
    \caption{\label{fig:threat_model}The considered threat model.}
\end{figure}

The threat model we consider consists of an attacker, a NIDS, and a supervised host, as illustrated in~\Cref{fig:threat_model}.
The NIDS gets a copy of all the network packets the host receives and sends.
We also suppose the NIDS is configured such that the supervised host's IP address is associated with the corresponding reassembly policy\footnote{Based on the current knowledge, we consider that hosts with more recent OS versions than the ones tested by Novak and Sturges in~\cite{novak2005target,novak2007target} (e.g., Debian 12) should have the latest OS family representative reassembly policy (e.g., \emph{linux} because Linux 2.4 was the latest tested version) associated in the NIDS configuration file.}, if the NIDS offers such a feature.
The attacker should be able to:

\begin{itemize}
    \item identify supervised host and NIDS reassembly policies, enabling them to choose a good (i.e., differently reassembled) overlap case candidate.
    \item craft IP header fields and payload to perform IP fragment-based attack.
    \item craft TCP header fields and payload to perform TCP segment-based attack.
\end{itemize}

\subsubsection*{Gaining supervised host and NIDS reassembly policy knowledge}

The attacker may want to learn about the monitored host and NIDS reassembly policies to increase the chances of a successful attack.
A good approximation to learn about the host reassembly policy is to determine the host OS.
Several tools exist to perform OS fingerprinting.
Active ones, such as Nmap~\cite{nmap} or Hershel(+)~\cite{shamsi2016unsupervised}, use specifically crafted packet sequences or retransmission times to identify a host OS, while passive tools, such as p0f \cite{p0f} or nPrintML \cite{holland2021new}, analyze packet header fields in existing communications.
Based on this knowledge and considering the hypothesis that the NIDS is correctly configured, the attacker can determine the NIDS reassembly policy.
Finally, they can craft an overlap chunk sequence based on the results reported in~\Cref{sec:ids_reassembly_consistency} and their objectives.

\subsubsection*{Crafting the data chunks}

According to the selected overlap case, the attacker chunks the original malicious packet into several pieces. 
They must appropriately manipulate the header fields \emph{Fragment Offset} and \emph{More Fragments} (resp. \emph{Sequence Number}) to perform an IP (resp. a TCP) chunk-based attack.
But, the other header fields of the crafted chunks must be consistent with the carried data (e.g., correct IP or TCP checksum, correct length).
The original payload must be in plaintext\footnote{The attacker may however take advantage of the overlap ambiguity during encryption initialization to corrupt the NIDS processing of TLS certificates or ciphersuites for example.}, and choosing the malicious, (i.e., "ATTACK" in~\Cref{tables/reassembly_cases_numbered}) payload depends on what the attacker wishes to do.
However, there are some constraints on the overlapping data portion which is \emph{not} visible in the non-target flow data, especially concerning the syntactic and semantic correctness of the upper-layer protocols.
This correctness depends on the performed attack scenario, i.e., \myhyperlink{E1}, \myhyperlink{E2}, \myhyperlink{I1}, or \myhyperlink{I2}, as described in~\Cref{tables/reassembly_cases_numbered}.

\paragraph{E1 and I1-related constraints}

The non-targeted implementation ignores the overlap chunk sequence; thus, no data is pushed to the upper protocol data stream.
It means that the overlapping data portion does not matter and, ultimately, \emph{any data} that fits the required length is, in fact, possible.
The overlapping data portions can even be the same.

\paragraph{E2 and I2-related constraints}

The NIDS and the supervised host both push data to the upper-layer protocol.
For the "AT00CK" reconstruction to be harmless, it must syntactically and semantically conform with all the upper protocols.
In particular, as for any upper-layer checksum:
1) if the checksum is contained within the overlapping portion, then it should be correctly adjusted to fit the "AT00CK" reassembled payload; otherwise,
2) the overlapping data should be adapted to fit the checksum\footnote{Only two dedicated octets are required to make a payload fit any internet checksum because it is computed with 2-octet words~\cite{rfc791}.
    See Appendix C "Deceiving TCP checksum" of Feng et al. work~\cite{feng2022pmtud} for a payload crafting example.}.
Any upper-layer syntax or semantic direspect may cause unwanted side effects (e.g., NIDS alerts, NIDS or supervised host failure) that could affect the attacker's stealthiness.

\section{Testing method}
\label{sec:method}

This section describes the overall method used to obtain OS and NIDS reassembly policies, which are then compared in order to find insertion and evasion opportunities.
First, it introduces the algebra used to model overlapping chunk sequences.
Then, the section describes all the test case characteristics.
Finally, it details the hosts we use to test NIDSes and OSes.

\subsection{Chunk sequence modeling}
\label{sec:chunk_sequence_modeling}

In the present subsection, we document how a spatio-temporal reasoning algebra can be adapted to model overlapping chunks.

\subsubsection{Spatio-temporal reasoning}
\label{sec:spatio_temporal_reasoning}

Spatio-temporal reasoning is particularly well suited to model packets of protocols that allow chunking and, thus, overlapping.
Indeed, we can associate byte offset with one
spatial dimension and arrival time with one temporal dimension.
Allen's interval algebra \cite{allen1983maintaining} is such a spatio-temporal algebra.
It consists of 13 different relations, which are described within the first two columns of~\Cref{tables/allen_relations}.
As we can see, there are four non-overlapping Allen relations, i.e., $M$, $Mi$, $B$, and $Bi$, and nine overlapping ones, i.e., $Eq$, $O$, $Oi$, $S$, $Si$, $D$, $Di$, $F$, and $Fi$.
The rightmost column transposes the relation meaning in terms of Internet packet sequence.

\renewcommand{\arraystretch}{1}
\begin{table}[t!]
    \setlength\tabcolsep{7pt}
    \centering
    \begin{tabular}{lc|cr|L{0.4\textwidth}}
        \toprule
        \multicolumn{2}{c|}{\textbf{\makecell{Relation                                                                                                                                                                                                                                                                                                                                                                                                    \\\textcolor{Red}{$\mathcal{R}$}}}}     &    \multicolumn{2}{c|}{\textbf{\makecell{Relation\\\textcolor{Red}{$\mathcal{R}$ i}nverse}}} & \multicolumn{1}{>{\centering\arraybackslash}m{0.4\textwidth}}{\textbf{Meaning}}                                                           \\
        \midrule
        X \textcolor{Red}{$M$} Y  & \raisebox{-9pt}{\includegraphics[width=0.1\textwidth]{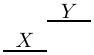}}  & X \textcolor{Red}{$Mi$} Y & \raisebox{-9pt}{\includegraphics[width=0.1\textwidth]{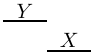}} & Meet: in-order (resp. out-of-order)  contiguous chunks                                                                                                              \\

        \hline

        X \textcolor{Red}{$B$} Y  & \raisebox{-8pt}{\includegraphics[width=0.1\textwidth]{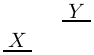}}  & X \textcolor{Red}{$Bi$} Y & \raisebox{-8pt}{\includegraphics[width=0.1\textwidth]{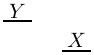}} & Before: data hole between one chunk       ending byte and the other chunk's payload starting byte                                                                   \\
        \hline
        \hline
        X \textcolor{Red}{$Eq$} Y & \raisebox{-8pt}{\includegraphics[width=0.1\textwidth]{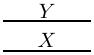}} & -                         &                                                                                                          & Equal: complete data overlap with         the                    chunks starting and finishing    at the same byte offsets. Data retransmissions are $Eq$ overlaps. \\
        \hline

        X \textcolor{Red}{$O$} Y  & \raisebox{-6pt}{\includegraphics[width=0.1\textwidth]{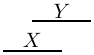}}  & X \textcolor{Red}{$Oi$} Y & \raisebox{-6pt}{\includegraphics[width=0.1\textwidth]{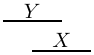}} & Overlap: partial data overlap                                                                                                                                       \\

        \hline

        X \textcolor{Red}{$S$} Y  & \raisebox{-6pt}{\includegraphics[width=0.1\textwidth]{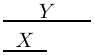}}  & X \textcolor{Red}{$Si$} Y & \raisebox{-6pt}{\includegraphics[width=0.1\textwidth]{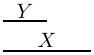}} & Start: partial data overlap                                                                                                                                         \\

        \hline

        X \textcolor{Red}{$D$} Y  & \raisebox{-6pt}{\includegraphics[width=0.1\textwidth]{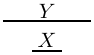}}  & X \textcolor{Red}{$Di$} Y & \raisebox{-6pt}{\includegraphics[width=0.1\textwidth]{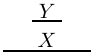}} & During: partial data overlap                                                                                                                                        \\

        \hline
        X \textcolor{Red}{$F$} Y  & \raisebox{-6pt}{\includegraphics[width=0.1\textwidth]{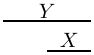}}  & X \textcolor{Red}{$Fi$} Y & \raisebox{-6pt}{\includegraphics[width=0.1\textwidth]{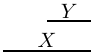}} & Finish: partial data overlap                                                                                                                                        \\

        \bottomrule
    \end{tabular}
    \caption{\label{tables/allen_relations}Allen's interval algebra relations and the corresponding meaning in terms of Internet packet sequences.}
\end{table}

While these relations describe the relative byte-wise and time-wise position of two chunks, they do not give any information regarding the chunk contents.
As a result, one cannot deduce from an overlapping relation whether the overlapping portion contains the same or different data.

\subsubsection{Overlap test case modeling}
\label{sec:test_case_modeling}

We use Allen's interval relations (or Allen relations for short) to ensure the exhaustiveness of overlap test cases.
In the following, a test case is always time-wisely described.
In other words, if $c_1 \ \mathcal{R} \ c_2$ with $c_1 \in C^p$ and $c_2 \in C^p$, then $t_1 < t_2$ (i.e., $c_1$ arrives before $c_2$).

There are nine overlapping Allen relations; thus, we consider that \emph{exhaustiveness in terms of overlap coverage is reached by testing these nine overlap cases}.
Thanks to the modeling, we can now prove that Novak and Sturges~\cite{novak2005target,novak2007target} and Atlasis~\cite{atlasis2012attacking} manually found and tested all the possible overlapping test cases.
See~\Cref{tables/state_of_the_art_summary} for the other work transposition into Allen formalism.

\subsection{Test case characteristics}
\label{sec:testing_characteristics}

\renewcommand{\arraystretch}{1}
\begin{table}[!t]
    \setlength\tabcolsep{4pt}
    \centering
    \begin{tabular}{ll|c}
        \toprule
        \textbf{\makecell[l]{Characteristic}} & \textbf{Description}                                       & \textbf{Example} \\
        \midrule
        Overlap type                          & Allen relation(s)                                          & \boldmath$O$     \\
        \cline{1-2}
        Chunk payload                         & \makecell[l]{AABBCCDD $\rightarrow$ DDCCBBAA                                  \\ensuring       checksum validity}    & $\downarrow$  \\
        \cline{1-2}
        \multirow{4}{*}{\makecell[l]{Reassembly                                                                               \\trigerring}} & \makecell[l]{\encircle{1} \emph{IP}: the rightmost finishing\\and lastly sent fragment has the\\\emph{More Fragments} (MF) bit unset} & \multirow{6}{*}{\includegraphics[width=0.35\textwidth]{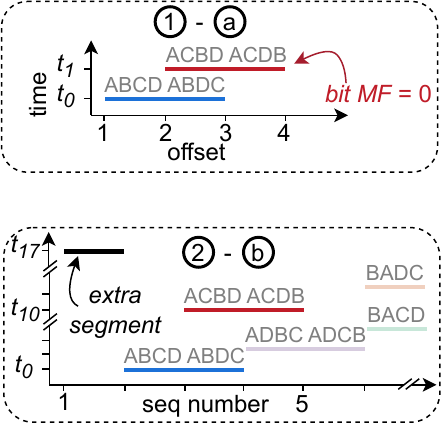}} \\
        \cline{2-2}
                                              & \makecell[l]{\encircle{2} \emph{TCP}: extra segment at the                    \\byte-wise                         beginning of the\\test case segments} \\
        \cline{1-2}
        \multirow{3}{*}{Mode}                 & \makecell[l]{\encircle{a} \emph{single}: overlaps tested                      \\individually} &                                           \\
        \cline{2-2}
                                              & \makecell[l]{\encircle{b} \emph{multiple}: overlaps tested                    \\altogether}                                           \\
        \cline{1-2}
        \multirow{2}{*}{\makecell[l]{Upper-layer                                                                              \\service}} & \emph{IP}: ICMP or ICMPv6 Echo \\
                                              & \emph{TCP}: TCP Echo                                                          \\
        \bottomrule
    \end{tabular}
    \caption{\label{tables/testing_characteristics} The IP and TCP test case characteristics.}
\end{table}

\Cref{tables/testing_characteristics} summarizes all the overlap test case characteristics.
The chunk payloads of a test case are chosen so that
1) they align with the IP header field's unit, which is 8-byte,
2) no matter which overlapping data is preferred, the higher layer checksum is valid, and, of course,
3) the preferred chunk can be distinguished from the other.
Novak introduced these payload patterns in~\cite{novak2005target}.
We also use Novak and Sturges's trick~\cite{novak2005target,novak2007target}, which ensures that all the chunks have been received before the chunk sequence reconstruction occurs, with \encircle{1} and \encircle{2}. 
Finally, overlaps can be tested $singly$ within nine separate chunk sequences \encircle{a} as~\cite{novak2007target,ptacek1998insertion} did ("individual overlap tests" in ~\cite{novak2007target}).
Or, differently, $multiple$ overlaps can be tested altogether within a unique chunk sequence, resulting in one reassembly \encircle{b} (Novak and Sturges name it "model overlap tests" in ~\cite{novak2007target}).
This last mode has been the most tested in the related works that targeted the OSes~\cite{di2023new,novak2005target,novak2007target,shankar2003active}.
We specifically use Novak and Sturges' \emph{multiple} chunk sequence.
The third column of~\Cref{tables/testing_characteristics} illustrates the $O$ relation's test for some introduced characteristics, with simplified chunk payloads to reduce figure size.

\subsection{OS and NIDS host targets}
\label{sec:os_nids_characteristics}

We perform OS testing through a classical Base-Target architecture.
The targeted OSes are varying Vagrant/Virtualbox-based boxes.
The testing scripts are all launched from the Base box.
As for the NIDSes, they are tested within Docker if an official image exists; otherwise, the tests are performed locally.
To deduce NIDS reassembly policies, we alert on 1) the chunk payload patterns introduced in~\Cref{tables/testing_characteristics} and 2) the upper-layer service (i.e., ICMP for IPv4, ICMPv6 for IPv6, and port 7 for TCP).
The following IPv4 entry rules are, for example, used to match on the "AABBCCDD" pattern:
\begin{itemize}
    \item Suricata and Snort: \textrulefont{alert$\; $icmp$\; $[192.168.0.1]$\; $any$\; $->$\; $any$\; $any$\: $(msg:\\"AABBCCDD$\; $detected";$\; $content:"AABBCCDD";$\; $sid:1;$\; $rev:7;)}
    \item Zeek\footnote{Since Zeek's preferred pattern-matching method is scripting, we verified that test case reassemblies are the same across the two matching methods.}: \textrulefont{signature$\; $ipv4-AABBCCDD$\; $\{$\; $ip-proto$\; $==$\; $icmp$\ $src-ip$\; $==\\192.168.0.1$\ $payload$\; $/.*AABBCCDD.*/$\ $event$\; $"AABBCCDD\\detected"\}}
\end{itemize}
In both cases, Network Interface Controller (NIC) offloading is disabled so as not to interfere with the targeted implementation reassembly.
See~\Cref{sec:chunk_alteration_before} for more details on OS and NIDS reassembly interferences.

\section{Results}
\label{sec:results}

In this section, we first describe IPv4, IPv6, and TCP reassembly policies of some OSes. The tested OS versions cover a large spectrum for the last 10 years. We then verify NIDS reassembly consistency with these OSes. The versions we target are: 
\begin{itemize}
    \item \emph{OS}: Windows 10 (21h2) and 11 (23h2), Debian 9 (Linux 4.9) and 12 (Linux 6.1), FreeBSD 10.2, 12.1 and 14.1, OpenBSD\footnote{The tested FreeBSD and OpenBSD OS versions reassemble the same way the overlap test cases; therefore, we only report FreeBSD policies.} 6.0, 6.9 and 7.6 and, Solaris 11.2 to 11.4 (SunOS 5.11). 
    \item \emph{NIDS}: Suricata v7.0.4, Snort v3.1.83, and Zeek v6.2.0.
\end{itemize}

\subsection{OS reassembly policies}
\label{sec:os_reassemblies}

This sub-section details IPv4, IPv6, and TCP reassembly policies for recent OSes in both \emph{multiple} and \emph{single} testing modes.
When relevant, we also compare our findings with the latest related work: IPv4 \emph{multiple} mode testing from~\cite{novak2005target} findings, IPv6 \emph{multiple} mode from~\cite{di2023new}, and TCP \emph{multiple} and \emph{single} modes from~\cite{novak2007target}.

\subsubsection{IP protocols}

\Cref{tables/ip_os_pair_pep} reports OS IPv4 and IPv6 test case reassembly policies.

\renewcommand{\arraystretch}{1}

\begin{table}[t!]
    \setlength\tabcolsep{4pt}
    \centering
    \small

    \begin{tabular}{c c c c c c c c c c c c}
        \toprule

        \multirow{3}{*}{\textbf{OS kernel}} & \multirow{3}{*}{\textbf{\makecell{Protocol                                                                                                                                                                                                                                                                                                                                                                                                                                                    \\version}}} &  \multicolumn{10}{c}{\textbf{Test case}}                       \\
        \cline{3-12}
                                     &                                            & \multirow{2}{*}{\textbf{\makecell{Testing                                                                                                                                                                                                                                                                                                                                                                                                        \\mode}}} & \multicolumn{9}{c}{\textbf{Overlapping relation}}                                                                                                                                                                                                                                                                                                                                                                                                                                                      \\
        \cline{4-12}
                                     &                                            &                                           & \boldmath$F$                         & \boldmath$Fi$                        & \boldmath$S$                         & \boldmath$Si$                        & \boldmath$O$                                   & \boldmath$Oi$                                  & \boldmath$D$                         & \boldmath$Di$                        & \boldmath$Eq$                                            \\
        \midrule
        \multirow{4}{*}{\makecell{Windows                                                                                                                                                                                                                                                                                                                                                                                                                                                                                            \\21h2, 23h2}}             & \multirow{2}{*}{v4}                    & \emph{multiple}                           & \cellcolor{red!15}\color{Blue}{\boldmath$\varnothing$} & \cellcolor{red!15}\color{Black}{$\varnothing$} & \cellcolor{red!15}\color{Blue}{\boldmath$\varnothing$} & \cellcolor{red!15}\color{Blue}{\boldmath$\varnothing$} & \cellcolor{red!15}\color{Black}{$\varnothing$} & \cellcolor{red!15}\color{Black}{$\varnothing$} & \cellcolor{red!15}\color{Blue}{\boldmath$\varnothing$} & \cellcolor{red!15}\color{Blue}{\boldmath$\varnothing$} & \cellcolor{red!15}\color{Blue}{\boldmath$\varnothing$} \\
                                     &                                            & \emph{single}                             & \color{Blue}{\textbf{n}}             & \color{Black}{$\varnothing$}         & \color{Blue}{\textbf{n}}             & \color{Blue}{\textbf{o}}             & \color{Black}{$\varnothing$}                   & \color{Black}{$\varnothing$}                   & \color{Blue}{\textbf{n}}             & \color{Blue}{\textbf{o}}             & \color{Blue}{\textbf{n}}                                 \\
        \cline{3-12}
                                     & \multirow{2}{*}{v6}                        & \emph{multiple}                           & \color{Blue}{\boldmath$\varnothing$} & \color{Black}{$\varnothing$}         & \color{Blue}{\boldmath$\varnothing$} & \color{Blue}{\boldmath$\varnothing$} & \cellcolor{red!15}\color{Black}{$\varnothing$} & \cellcolor{red!15}\color{Black}{$\varnothing$} & \color{Blue}{\boldmath$\varnothing$} & \color{Blue}{\boldmath$\varnothing$} & \cellcolor{red!15}\color{Blue}{\boldmath$\varnothing$}   \\
                                     &                                            & \emph{single}                             & \color{Blue}{\textbf{n}}             & \color{Black}{$\varnothing$}         & \color{Blue}{\textbf{n}}             & \color{Blue}{\textbf{o}}             & \color{Black}{$\varnothing$}                   & \color{Black}{$\varnothing$}                   & \color{Blue}{\textbf{n}}             & \color{Blue}{\textbf{o}}             & \color{Blue}{\textbf{n}}                                 \\

        \hline
        \multirow{4}{*}{\makecell{Linux                                                                                                                                                                                                                                                                                                                                                                                                                                                                                             \\4.9, 6.1}}              & \multirow{2}{*}{v4}                    & \emph{multiple}                           & \cellcolor{red!15}\color{Blue}{\boldmath$\varnothing$} & \cellcolor{red!15}\color{Black}{$\varnothing$} & \cellcolor{red!15}\color{Blue}{\boldmath$\varnothing$} & \cellcolor{red!15}\color{Blue}{\boldmath$\varnothing$} & \cellcolor{red!15}\color{Black}{$\varnothing$} & \cellcolor{red!15}\color{Black}{$\varnothing$} & \cellcolor{red!15}\color{Blue}{\boldmath$\varnothing$} & \cellcolor{red!15}\color{Blue}{\boldmath$\varnothing$} & \cellcolor{red!15}\color{Blue}{\boldmath$\varnothing$} \\
                                     &                                            & \emph{single}                             & \color{Blue}{\textbf{n}}             & \color{Black}{$\varnothing$}         & \color{Blue}{\textbf{n}}             & \color{Blue}{\textbf{o}}             & \color{Black}{$\varnothing$}                   & \color{Black}{$\varnothing$}                   & \color{Blue}{\textbf{n}}             & \color{Blue}{\textbf{o}}             & \color{Blue}{\textbf{n}}                                 \\
        \cline{3-12}
                                     & \multirow{2}{*}{v6}                        & \emph{multiple}                           & \color{Blue}{\boldmath$\varnothing$} & \color{Black}{$\varnothing$}         & \color{Blue}{\boldmath$\varnothing$} & \color{Blue}{\boldmath$\varnothing$} & \cellcolor{red!15}\color{Black}{$\varnothing$} & \cellcolor{red!15}\color{Black}{$\varnothing$} & \color{Blue}{\boldmath$\varnothing$} & \color{Blue}{\boldmath$\varnothing$} & \cellcolor{red!15}\color{Blue}{\boldmath$\varnothing$}   \\
                                     &                                            & \emph{single}                             & \color{Blue}{\textbf{n}}             & \color{Black}{$\varnothing$}         & \color{Blue}{\textbf{n}}             & \color{Blue}{\textbf{o}}             & \color{Black}{$\varnothing$}                   & \color{Black}{$\varnothing$}                   & \color{Blue}{\textbf{n}}             & \color{Blue}{\textbf{o}}             & \color{Blue}{\textbf{n}}                                 \\

        \hline
        \multirow{4}{*}{\makecell{SunOS                                                                                                                                                                                                                                                                                                                                                                                                                                                                                              \\5.11}}               & \multirow{2}{*}{v4}                    & \emph{multiple}                           & \cellcolor{Green!10}\color{Black}{n}                   & \cellcolor{Green!10}\color{Blue}{\textbf{o}}   & \cellcolor{Green!10}\color{Blue}{\textbf{o}}           & \cellcolor{Green!10}\color{Black}{o}                   & \cellcolor{Green!10}\color{Black}{o}           & \cellcolor{Green!10}\color{Black}{o}           & \cellcolor{Green!10}\color{Black}{n}                   & \cellcolor{Green!10}\color{Black}{o}                   & \cellcolor{Green!10}\color{Blue}{\textbf{o}}           \\
                                     &                                            & \emph{single}                             & \color{Black}{n}                     & \color{Blue}{\boldmath$\varnothing$} & \color{Blue}{\textbf{n}}             & \color{Black}{o}                     & \color{Black}{o}                               & \color{Black}{o}                               & \color{Black}{n}                     & \color{Black}{o}                     & \color{Blue}{\textbf{n}}                                 \\
        \cline{3-12}
                                     & \multirow{2}{*}{v6}                        & \emph{multiple}                           & \color{Black}{n}                     & \color{Blue}{\textbf{o}}             & \color{Blue}{\textbf{o}}             & \color{Black}{o}                     & \color{Black}{o}                               & \color{Black}{o}                               & \color{Black}{n}                     & \color{Black}{o}                     & \color{Blue}{\textbf{o}}                                 \\
                                     &                                            & \emph{single}                             & \color{Black}{n}                     & \color{Blue}{\boldmath$\varnothing$} & \color{Blue}{\textbf{n}}             & \color{Black}{o}                     & \color{Black}{o}                               & \color{Black}{o}                               & \color{Black}{n}                     & \color{Black}{o}                     & \color{Blue}{\textbf{n}}                                 \\

        \hline
        \multirow{4}{*}{\makecell{FreeBSD                                                                                                                                                                                                                                                                                                                                                                                                                                                                                            \\10.2, 12.1, 14.2}}   & \multirow{2}{*}{v4}        & \emph{multiple} & \cellcolor{Green!10}\color{Black}{n}           & \cellcolor{Green!10}\color{Blue}{\textbf{o}}               & \cellcolor{Green!10}\color{Blue}{\textbf{o}}             & \cellcolor{Green!10}\color{Black}{o}           & \cellcolor{Green!10}\color{Black}{o}               & \cellcolor{Green!10}\color{Black}{n}               & \cellcolor{Green!10}\color{Black}{n}           & \cellcolor{Green!10}\color{Black}{o}           & \cellcolor{Green!10}\color{Blue}{\textbf{o}}             \\
                                     &                                            & \emph{single}                             & \color{Black}{n}                     & \color{Blue}{\boldmath$\varnothing$} & \color{Blue}{\textbf{n}}             & \color{Black}{o}                     & \color{Black}{o}                               & \color{Black}{n}                               & \color{Black}{n}                     & \color{Black}{o}                     & \color{Blue}{\textbf{n}}                                 \\
        \cline{3-12}
                                     & \multirow{2}{*}{v6}                        & \emph{multiple}                           & \color{Blue}{\boldmath$\varnothing$} & $\varnothing$                        & \color{Blue}{\boldmath$\varnothing$} & \color{Blue}{\boldmath$\varnothing$} & \cellcolor{Green!10}$\varnothing$              & \cellcolor{Green!10}$\varnothing$              & \color{Blue}{\boldmath$\varnothing$} & \color{Blue}{\boldmath$\varnothing$} & \cellcolor{Green!10}\color{Blue}{\boldmath$\varnothing$} \\
                                     &                                            & \emph{single}                             & \color{Blue}{\textbf{n}}             & $\varnothing$                        & \color{Blue}{\textbf{n}}             & \color{Blue}{\textbf{o}}             & $\varnothing$                                  & $\varnothing$                                  & \color{Blue}{\textbf{n}}             & \color{Blue}{\textbf{o}}             & \color{Blue}{\textbf{n}}                                 \\

        \bottomrule
    \end{tabular}
    \caption{\label{tables/ip_os_pair_pep}OS IP reassembly policies.
        o (resp. n) means that oldest (resp. newest) chunk data is prefered and $\varnothing$ means that the OS ignores the overlap.
        \textcolor{Blue}{Bold blue} means that testing modes are reassembled differently.
        IPv4 (resp. IPv6) cell backgrounds encodes \colorbox{red!15}{in}\colorbox{Green!10}{consistency} with \cite{novak2005target} (resp. \cite{di2023new}).
    }
\end{table}

\paragraph*{IPv4}

Windows and Linux policies have evolved for the \emph{multiple} testing mode since Novak~\cite{novak2005target}.
Both OSes now ignore overlap chunks.
The other OSes have not changed their policies.
However, the newly tested mode, namely \emph{single}, shows different reassemblies for all the OSes when compared to the \emph{multiple} mode.
In total, 6 out of 9 overlapping relations are reassembled differently for Windows and Linux-based OSes,
while it accounts for 3 out of 9 for the remaining OSes.
We hypothesize that the context introduced by the adjacent chunks used inside \emph{multiple} mode causes the discrepancies between the two testing modes.
We thus argue that the \emph{single} mode reassemblies should be used to obtain context-agnostic reassemblies inside the NIDSes.
In this testing mode, all the OSes reassemble $F$, $S$, $Si$, $D$, $Di$, and $Eq$ relations the same way, never ignoring the test cases.
$Fi$ relation is never reassembled, possibly due to the fragment with the \emph{MF} bit unset's drop.
Finally, $O$ and $Oi$ are the only overlap test cases that show different reassemblies depending on the OS.

\paragraph*{IPv6} FreeBSD OSes do not reassemble $O$ and $Oi$ Allen relations, which differs from the observed IPv4 behavior.
Except for FreeBSD, all OSes reassemble IPv4 and IPv6 overlapping fragments the same way, and thus, the previous paragraph descriptions also apply to IPv6 fragments.
The Windows and Linux-based OSes ignore all the overlapping relations in a \emph{multiple} test mode, which is inconsistent with Di Paolo's~\cite{di2023new} findings for $O$, $Oi$, and $Eq$ relations.
Since the OS versions that Di Paolo tested are very close to ours, we hypothesize that the lack of tested relation set exhaustiveness
for that mode impacts the extracted reassembly.

\subsubsection{TCP protocol}

\Cref{tables/tcp_os_pair_peosf} describes OS TCP reassembly policies and compares findings with state-of-the-art ones ~\cite{novak2007target}.
Windows and FreeBSD reassemble similarly the overlaps across testing modes.
The former OS always reassembles with the oldest segment data.
These policies are consistent with~\cite{novak2007target} description.
The latest Linux-based OSes reassemble $S$ relation differently depending on the testing mode, favoring old (resp. new) data for \emph{multiple} (resp. \emph{single}) mode.
SunOS 5.11 also reassembles the $Eq$ overlap differently, which is consistent with \cite{novak2007target} findings.

\renewcommand{\arraystretch}{1}

\begin{table}[t!]
    \setlength\tabcolsep{5pt}
    \centering
    \small

    \begin{tabular}{c c c c c c c c c c c}
        \toprule

        \multirow{3}{*}{\textbf{OS kernel}} & \multicolumn{9}{c}{\textbf{Test case}}                                                                                                                                                                                                                                                                                                                                                                                                                                                                                                                                     \\
        \cline{2-11}
                                     & \multirow{2}{*}{\textbf{\makecell{Testing                                                                                                                                                                                                                                                                                                                                                                                                                                                                                                                                  \\mode}}} & \multicolumn{9}{c}{\textbf{Overlapping relation}}                                                                                                                                                                                                                                                                                                                                                                                                                                                      \\
        \cline{3-11}
                                     &                                           & \boldmath$F$                                           & \boldmath$Fi$                                          & \boldmath$S$                                           & \boldmath$Si$                                          & \boldmath$O$                                           & \boldmath$Oi$                                          & \boldmath$D$                                           & \boldmath$Di$                                          & \boldmath$Eq$                                          \\

        \midrule
        \makecell{Windows                                                                                                                                                                                                                                                                                                                                                                                                                                                                                                                                                                                         \\21h2, 23h2}   & \emph{any}                                             & \cellcolor{Green!10}\color{Black}{o} & \cellcolor{Green!10}\color{Black}{o} & \cellcolor{Green!10}\color{Black}{o} & \cellcolor{Green!10}\color{Black}{o} & \cellcolor{Green!10}\color{Black}{o} & \cellcolor{Green!10}\color{Black}{o} & \cellcolor{Green!10}\color{Black}{o} & \cellcolor{Green!10}\color{Black}{o} & \cellcolor{Green!10}\color{Black}{o} \\

        \hline
        Linux                       & \emph{multiple}                           & \cellcolor{Green!10}\color{Black}{n}                   & \cellcolor{Green!10}\color{Black}{o}                   & \cellcolor{Red!15}\color{Blue}{\textbf{o}}             & \cellcolor{Green!10}\color{Black}{o}                   & \cellcolor{Green!10}\color{Black}{o}                   & \cellcolor{Green!10}\color{Black}{n}                   & \cellcolor{Green!10}\color{Black}{n}                   & \cellcolor{Green!10}\color{Black}{o}                   & \cellcolor{Green!10}\color{Black}{o}                   \\
        4.9, 6.1                           & \emph{single}                             & \cellcolor{Green!10}\color{Black}{n}                   & \cellcolor{Green!10}\color{Black}{o}                   & \cellcolor{Green!10}\color{Blue}{\textbf{n}}           & \cellcolor{Green!10}\color{Black}{o}                   & \cellcolor{Green!10}\color{Black}{o}                   & \cellcolor{Green!10}\color{Black}{n}                   & \cellcolor{Green!10}\color{Black}{n}                   & \cellcolor{Green!10}\color{Black}{o}                   & \cellcolor{Green!10}\color{Black}{o}                   \\

        \hline
        SunOS                        & \emph{multiple}                           & \cellcolor{Green!10}\color{Black}{n}                   & \cellcolor{Green!10}\color{Black}{o}                   & \cellcolor{Green!10}\color{Black}{n}                   & \cellcolor{Green!10}\color{Black}{o}                   & \cellcolor{Green!10}\color{Black}{n}                   & \cellcolor{Green!10}\color{Black}{o}                   & \cellcolor{Green!10}\color{Black}{n}                   & \cellcolor{Green!10}\color{Black}{o}                   & \cellcolor{Green!10}\color{Blue}{\textbf{n}}           \\
        5.11                         & \emph{single}                             & \cellcolor{Green!10}\color{Black}{n}                   & \cellcolor{Green!10}\color{Black}{o}                   & \cellcolor{Green!10}\color{Black}{n}                   & \cellcolor{Green!10}\color{Black}{o}                   & \cellcolor{Green!10}\color{Black}{n}                   & \cellcolor{Green!10}\color{Black}{o}                   & \cellcolor{Green!10}\color{Black}{n}                   & \cellcolor{Green!10}\color{Black}{o}                   & \cellcolor{Green!10}\color{Blue}{\textbf{o}}           \\

        \hline
        FreeBSD                      & \multirow{2}{*}{\emph{any}}               & \cellcolor{Green!10}                                   & \cellcolor{Green!10}                                   & \cellcolor{Green!10}                                   & \cellcolor{Green!10}                                   & \cellcolor{Green!10}                                   & \cellcolor{Green!10}                                   & \cellcolor{Green!10}                                   & \cellcolor{Green!10}                                   & \cellcolor{Green!10}                                   \\
        10.2, 12.1, 14.2                         &                                           & \cellcolor{Green!10}\multirow{-2}{*}{\color{Black}{n}} & \cellcolor{Green!10}\multirow{-2}{*}{\color{Black}{o}} & \cellcolor{Green!10}\multirow{-2}{*}{\color{Black}{o}} & \cellcolor{Green!10}\multirow{-2}{*}{\color{Black}{o}} & \cellcolor{Green!10}\multirow{-2}{*}{\color{Black}{o}} & \cellcolor{Green!10}\multirow{-2}{*}{\color{Black}{n}} & \cellcolor{Green!10}\multirow{-2}{*}{\color{Black}{n}} & \cellcolor{Green!10}\multirow{-2}{*}{\color{Black}{o}} & \cellcolor{Green!10}\multirow{-2}{*}{\color{Black}{o}} \\

        \bottomrule
    \end{tabular}
    \caption{\label{tables/tcp_os_pair_peosf}OS TCP reassembly policies.
        o (resp. n) means that oldest (resp. newest) chunk data is prefered.
        \textcolor{Blue}{Bold blue} means that testing modes are reassembled differently.
        Cell backgrounds encodes \colorbox{red!15}{in}\colorbox{Green!10}{consistency} with \cite{novak2007target}.
    }

\end{table}

\subsubsection{Takeaways}
\label{sec:os_reassembly_takeaways}

IPv4, IPv6, and TCP reassembly policies are more complex than described in the state-of-the-art as overlap reassemblies change depending on the test mode.
IP policies have evolved; for example, the Windows and Linux families now show the same reassemblies.
TCP reassembly policies have been unchanged since 2007, except Linux's.
Finally, OSes continue to reassemble some overlap test cases differently by favoring old or new data or ignoring the chunks.

\subsection{NIDS/OS reassembly consistency}
\label{sec:ids_reassembly_consistency}

This section compares the NIDS reassembly policies we observed with the ones of OSes (see \Cref{sec:os_reassemblies}) for IPv4, IPv6, and TCP protocols and the \emph{single} mode.
Because of space issues, we first check reassembly discrepancies between NIDSes and one OS family, namely FreeBSD.
We choose this OS because it offers the most insertion and evasion attack\footnote{Based on~\Cref{sec:problem_defintion} definitions, if the NIDS and the host reassemble differently a test case by not ignoring it (i.e., \myhyperlink{E2} and \myhyperlink{I2} from~\Cref{tables/reassembly_cases_numbered}), the attacker can either perfom an insertion or an evasion. The other attack cases \myhyperlink{E1} and \myhyperlink{I1} are straightforward.} opportunities when configured inside Snort and Suricata using the \emph{bsd} policy.
We finish by summarizing results for all the tested OSes and providing metrics on NIDS attack opportunities. 
Full results can be found in \url{https://gitlab.inria.fr/laubard/dimva\_2025\_artifacts}.  

\subsubsection{Consistency with FreeBSD OSes}

\Cref{tables/freebsd-ids-pair-consistency} gathers IP and TCP NIDS reassembly policies in the \emph{single} mode, which is the easiest an attacker can exploit.

\renewcommand{\arraystretch}{1}

\begin{table}[t!]
    \setlength\tabcolsep{3pt}
    \centering
    \small

    \begin{tabular}{c l c c c c c c c c c}
        \toprule

        \multirow{2}{*}{\textbf{Protocol}} & \multirow{2}{*}{\textbf{Implementation}} & \multicolumn{9}{c}{\textbf{Overlapping relation}}                                                                                                                                                                                                     \\
        \cline{3-11}
                                           &                                          & \boldmath$F$                                      & \boldmath$Fi$                    & \boldmath$S$         & \boldmath$Si$        & \boldmath$O$         & \boldmath$Oi$        & \boldmath$D$         & \boldmath$Di$        & \boldmath$Eq$        \\
        \midrule
        \multirow{4}{*}{IPv4}              & \textbf{FreeBSD}                    & \textbf{n}                                        & $\varnothing$                    & \textbf{n}           & \textbf{o}           & \textbf{o}           & \textbf{n}           & \textbf{n}           & \textbf{o}           & \textbf{n}           \\
                                           & Suricata-\emph{bsd}                      & \textcolor{Green}{n}                              & \textcolor{Red}{o}               & \textcolor{Green}{n} & \textcolor{Green}{o} & \textcolor{Green}{o} & \textcolor{Red}{o}   & \textcolor{Green}{n} & \textcolor{Green}{o} & \textcolor{Green}{n} \\
                                           & Snort-\emph{bsd}                         & \color{Green}{n}                                  & \textcolor{Green}{$\varnothing$} & \color{Green}{n}     & \color{Green}{o}     & \color{Green}{o}     & \color{Green}{n}     & \color{Green}{n}     & \color{Green}{o}     & \color{Green}{n}     \\
                                           & Zeek                                     & \textcolor{Green}{n}                              & \color{Red}{o}                   & \textcolor{Green}{n} & \textcolor{Green}{o} & \color{Green}{o}     & \color{Red}{o}       & \textcolor{Green}{n} & \textcolor{Green}{o} & \textcolor{Green}{n} \\

        \hline
        \multirow{4}{*}{IPv6}              & \textbf{FreeBSD}                    & \textbf{n}                                        & $\varnothing$                    & \textbf{n}           & \textbf{o}           & $\varnothing$        & $\varnothing$        & \textbf{n}           & \textbf{o}           & \textbf{n}           \\
                                           & Suricata-\emph{bsd}                      & \textcolor{Green}{n}                              & \textcolor{Red}{o}               & \textcolor{Green}{n} & \textcolor{Green}{o} & \textcolor{Red}{o}   & \textcolor{Red}{o}   & \textcolor{Green}{n} & \textcolor{Green}{o} & \textcolor{Green}{n} \\
                                           & Snort-\emph{bsd}                         & \color{Green}{n}                                  & \textcolor{Green}{$\varnothing$} & \color{Green}{n}     & \color{Green}{o}     & \color{Red}{o}       & \color{Red}{n}       & \color{Green}{n}     & \color{Green}{o}     & \color{Green}{n}     \\
                                           & Zeek                                     & \textcolor{Red}{o}                                & \textcolor{Red}{o}               & \textcolor{Red}{o}   & \textcolor{Green}{o} & \textcolor{Red}{o}   & \textcolor{Red}{o}   & \textcolor{Red}{o}   & \textcolor{Green}{o} & \textcolor{Red}{o}   \\

        \hline
        \multirow{5}{*}{TCP}               & \textbf{FreeBSD}                    & \textbf{n}                                        & \textbf{o}                       & \textbf{o}           & \textbf{o}           & \textbf{o}           & \textbf{n}           & \textbf{n}           & \textbf{o}           & \textbf{o}           \\
                                           & Snort-\emph{bsd}                         & \textcolor{Green}{n}                              & \textcolor{Green}{o}             & \textcolor{Green}{o} & \textcolor{Green}{o} & \textcolor{Green}{o} & \textcolor{Green}{n} & \textcolor{Green}{n} & \textcolor{Green}{o} & \textcolor{Green}{o} \\
                                           & Suricata-\emph{bsd}                      & \textcolor{Green}{n}                              & \textcolor{Green}{o}             & \textcolor{Green}{o} & \textcolor{Green}{o} & \textcolor{Green}{o} & \textcolor{Green}{n} & \textcolor{Green}{n} & \textcolor{Green}{o} & \textcolor{Green}{o} \\
                                           & Zeek                                     & \textcolor{Red}{o}                                & \textcolor{Green}{o}             & \textcolor{Green}{o} & \textcolor{Green}{o} & \textcolor{Green}{o} & \textcolor{Red}{o}   & \textcolor{Red}{o}   & \textcolor{Green}{o} & \textcolor{Green}{o} \\

        \bottomrule
    \end{tabular}
    \caption{\label{tables/freebsd-ids-pair-consistency}NIDS reassembly consistency with FreeBSD 10.2, 12.1, 14.1 in a \emph{single} testing mode.
        o (resp. n) means that oldest (resp. newest) chunk data is prefered and $\varnothing$ means the OS ignores the test case.
        \textcolor{Green}{Green} (resp. \textcolor{Red}{red}) means that NIDS reassembly is the same as (resp.  different from) FreeBSD.
    }
\end{table}

\paragraph*{IP}

NIDSes are all vulnerable to overlap-based attacks with at least two overlap types, except for Snort with IPv4 chunks.
Zeek and Suricata can be subject to IPv4 insertion attack with $Fi$ relation and IPv4 evasion or insertion with just $Oi$.
Moreover, with only two consistent IPv6 test case reassemblies with the FreeBSD, several more relations can be used to perform insertion or evasion attacks on Zeek.
Despite being based on the same \emph{bsd} policy description as~\cite{novak2005target}, Suricata reassembles IPv4 overlaps differently.
Snort is, interestingly, perfectly consistent with the OS for IPv4 protocol even though no previous work had described reassemblies for the \emph{single} test mode.
Additionaly, IPv4 and IPv6 fragments are reassembled similarly by Snort and Suricata.
Zeek notably reassembles all the overlaps by favoring the oldest IPv6 fragment data.

\paragraph*{TCP}

Snort and Suricata policies are consistent with the tested FreeBSD OSes for all the overlapping test cases.
The NIDSes, thus, consistently implemented the reassembly policies that Novak and Sturges described~\cite{novak2007target}.
Zeek, which does not have such a reassembly policy configuration capability, reassembles $F$, $Oi$, and $D$ inconsistently, as it always reassembles with the oldest segment's data.
An attacker can use these overlaps to perform insertion and evasion.

\subsubsection{Consistency with all OSes}

Snort reassembles perfectly consistently IPv4 test cases with the BSD and Sun-based OSes and IPv6 test cases with SunOS.
We can find at least one IP test case that is reassembled differently for the remaining OSes, i.e., the Linux and Windows ones.
Neither Suricata nor Zeek consistently reassembles all IP overlapping test cases with any of the characterized OSes.
Therefore, at least an insertion or an evasion attack can target these NIDS-OS couples.
Suppose that Snort or Suricata TCP reassembly policy is correctly associated with the host; there is no possible overlap-based attack except in one case: TCP \emph{solaris} policies and SunOS-based OSes with the $Eq$ test case.
On the contrary, Zeek reassembles segments consistently with Windows OSes but does not with the remaining OSes.
See \url{https://gitlab.inria.fr/laubard/dimva\_2025\_artifacts} for more details.

\renewcommand{\arraystretch}{1}

\begin{table}[t!]
    \setlength\tabcolsep{10pt}
    \centering
    \small

    \begin{tabular}{c l rr ll}
        \toprule

        \multirow{2}{*}{\textbf{Protocol}} & \multirow{2}{*}{\textbf{NIDS}} & \multicolumn{2}{c}{\multirow{2}{*}{\textbf{\makecell{Test case                                                                              \\inconsistencies}}}} & \multicolumn{2}{c}{\textbf{\makecell{Tested OS kernels with\\possible attack}}}             \\
        \cline{5-6}
                                           &                               &                                                                &      & \textbf{Evasion}                 & \textbf{Insertion}               \\

        \midrule
        \multirow{3}{*}{IPv4}              & Suricata                      & 8                                                              & 22\% & F & W, L, S, F    \\
                                           & Snort                         & 4                                                              & 11\% &   & W, L \\
                                           & Zeek                          & 9                                                              & 25\% & F    &  W, L, S, F  \\
        \hline
        \multirow{3}{*}{IPv6}              & Suricata                      & 9                                                              & 25\% &   & W, L, S, F    \\
                                           & Snort                         & 6                                                              & 17\% &   & W, L, F \\
                                           & Zeek                          & 28                                                             & 78\% & W, L, S, F    & W, L, S, F    \\
        \hline
        \multirow{3}{*}{TCP}               & Suricata                      & 1                                                              & 3\%  & S & S \\
                                           & Snort                         & 1                                                              & 3\%  & S & S \\
                                           & Zeek                          & 11                                                             & 31\% & L, S, F & L, S, F \\

        \bottomrule
    \end{tabular}
    \caption{\label{tables/ids_insertion_evasion_corresponding_reassembly_policies}NIDS inconsistencies with OS reassemblies and corresponding attack opportunities for a \emph{single} testing mode.  
        W, L, S, and F respectively correspond to the tested Windows, Linux, SunOS and, FreeBSD/OpenBSD kernels.
}
\end{table}

\Cref{tables/ids_insertion_evasion_corresponding_reassembly_policies} gives more general consistency metrics and related attack opportunities.
In particular, Zeek and Suricata reassemble IPv4 overlaps inconsistently for about $20\%$ of them.
Snort performs better with 11\% inconsistent test cases.
Snort and Suricata globally perform better than Zeek for IPv6, with the same or fewer OSes that can be targeted with an insertion or evasion attack.
Zeek, which has different IPv4 and IPv6 reassembly policies, performs worse for version 6 (28 inconsistencies) than for version 4 (9 inconsistencies).
TCP enables fewer attack opportunities, with only one overlap that Suricata and Snort reassemble incorrectly.
Zeek exhibits TCP-based evasion or insertion attacks for 3 OSes out of 4.
Since one inconsistency is enough to perform an insertion and/or an evasion, the OSes (i.e., columns 4 and 5 in~\Cref{tables/ids_insertion_evasion_corresponding_reassembly_policies}) that can be targeted are of importance. 

\subsubsection{Takeaways}
\label{sec:ids_reassembly_takeaways}

As expected, Snort and Suricata (which allow policy configuration) perform overall better than Zeek (which uses a unique policy).
Snort can protect itself completely against IPv4 and IPv6 evasion attacks (for the tested OSes) and almost entirely against TCP segment-based attacks.
Because OS reassembly policies evolve and are more complex than initially thought, NIDSes must (continuously) verify the consistency of the implemented policies.

\section{Related Works}
\label{sec:related_works}

This section presents the related works that analyzed and described IP and TCP implementation reassembly policies.
To ease the comparison with these works, we transpose the covered test cases into Allen's formalism in~\Cref{tables/state_of_the_art_summary}.
We categorize the works that tested implementation reassembly policies in two families according to the test case generation approach.

\renewcommand{\arraystretch}{1}

\begin{table}[t!]
    \setlength\tabcolsep{2.5pt}
    \centering
    \begin{tabular}{lll ccc c}
        \toprule
        \textbf{Author}                                  & \textbf{Work}                             & \textbf{Year}         & \textbf{Protocol}            & \textbf{\makecell{Testing                                                                \\mode}} & \textbf{\makecell{Tested\\Allen relations}} & \textbf{\makecell{Target\\type}} \\
        \midrule
        \makecell[l]{Ptacek$\:$et$\:$al.}                & \cite{ptacek1998insertion}                & 1998                  & IPv4/TCP                     & \emph{single}             & $Fi$, $D$                              & NIDS/OS              \\
        \hline
        \multirow{2}{*}{\makecell{Shankar$\:$et$\:$al.}} & \multirow{2}{*}{\cite{shankar2003active}} & \multirow{2}{*}{2003} & IPv4                         & \emph{multiple}           & $O$, $Oi$, $Eq$                        & \multirow{2}{*}{OS} \\
        \cline{4-6}
                                                         &                                           &                       & TCP                          & \emph{multiple}           & $O$, $D$                                                     \\
        \hline
        \multirow{3}{*}{\makecell{Novak$\:$et$\:$al.}}   & \cite{novak2005target}                    & 2005                  & IPv4                         & \emph{multiple}           & \multirow{3}{*}{\textcolor{blue}{all}} & \multirow{3}{*}{OS} \\
        \cline{2-5}
                                                         & \multirow{2}{*}{\cite{novak2007target}}   & \multirow{2}{*}{2007} & \multirow{2}{*}{TCP}         & \emph{multiple}           &                                                              \\
                                                         &                                           &                       &                              & \emph{single}             &                                                              \\
        \hline
        Atlasis                                          & \cite{atlasis2012attacking}               & 2012                  & IPv6                         & \emph{Na}                 & \textcolor{blue}{all}                  & OS                  \\
        \hline
        Khattak$\:$et$\:$al.                             & \cite{khattak2013towards}                 & 2013                  & IPv4/TCP                     & \emph{single}             & \textcolor{blue}{all}                  & CS                  \\
        \hline
        Wang$\:$et$\:$al.                                & \cite{wang2017your}                       & 2017                  & IPv4/TCP                     & \emph{single}             & $Eq$                                   & CS                  \\
        \hline
        Lin$\:$et$\:$al.                                 & \cite{lin2024research}                    & 2024                  & IPv6                         & \emph{single}             & $Eq$                                   & NIDS/OS              \\
        \midrule
        \midrule
        Bock$\:$et$\:$al.                                & \cite{bock2019geneva}                     & 2019                  & IPv4/TCP                     & \emph{Na}                 & Unknown                                & CS                  \\
        \hline
        \multirow{2}{*}{Wang$\:$et$\:$al.}               & \cite{wang2020symtcp}                     & 2020                  & TCP                          & \emph{Na}                 & \makecell{$F$, $D$, $Oi$}              & CS/NIDS              \\
        \cline{2-7}
                                                         & \cite{wang2021themis}                     & 2021                  & TCP                          & \emph{Na}                 & -                                      & OS                  \\
        \hline
        Zhang$\:$et$\:$al.                               & \cite{zhang2022statediver}                & 2022                  & IPv4/TCP                     & \emph{Na}                 & \makecell{$Eq$/Unknown}                & NIDS                 \\
        \hline
        Di Paolo$\:$et$\:$al.                            & \cite{di2023new}                          & 2023                  & IPv6                         & \emph{multiple}           & \makecell{$O$, $Oi$, $Eq$}             & OS                  \\
        \midrule
        \midrule
        \textbf{Us}                                      & \textbf{-}                                & \textbf{-}            & \textbf{\makecell{IPv4/IPv6/                                                                                            \\TCP}} & \textbf{\makecell{\emph{multiple}                                         \\ {\emph{single}}}} & \textbf{\textcolor{blue}{all}} & \textbf{NIDS/OS}\\
        \bottomrule
    \end{tabular}
    \caption{\label{tables/state_of_the_art_summary}Summary regarding overlap-based works.
        "Unknown" means that there is partial or no information on the covered relations for the work tool's run.
    }
\end{table}

\subsection{Manually generated overlap cases}
\label{sec:manual_related_works}

In 1998, Ptacek and Newsham~\cite{ptacek1998insertion} first showed that OSes could behave differently when reconstructing overlapping IPv4 and TCP chunks.
The reassembly ambiguity it poses for NIDSes opened a new research axis, and several works~\cite{shankar2003active,novak2005target,novak2007target,atlasis2012attacking} tried to unveil the IPv4, IPv6, and TCP reassembly policies of OSes.
Novak and Sturges's~\cite{novak2005target,novak2007target} works reached exhaustivity for the first time regarding the tested overlap types.
More recently, Lin et al.~\cite{lin2024research} showed that some OS and NIDS do not comply with RFC~\cite{deering2017rfc} as regards IPv6 data overlaps (which states that implementations should discard the entire fragment sequence in the presence of any overlap type).
However, since they tested a unique overlap type, they may have overestimated OS compliance.

In parallel, other works tried to evade censorship systems (CS) with different elusive packet sequence strategies.
In 2013, Khattak et al.~\cite{khattak2013towards} described the IPv4 and TCP reassembly policies of the Great Firewall of China (GFW) based on the complete set of overlap relations.
Wang et al.~\cite{wang2017your}, which manually designed a unique IPv4 and TCP overlap test case, showed that some middleboxes could interfere with the (original) overlapping fragment sequence by dropping or reassembling it, eliminating, thus, any ambiguity.
These works, however, do not consider that the server OSes may have different reassembly policies.

Overall, only 3 of the 8 works were exhaustive in regards to the covered overlap relations.
The lack of a unified overlap formalization may explain why most recent works target fewer overlap types than before, as shown in~\Cref{tables/state_of_the_art_summary}.
Finally, none of the works conducted a complete reassembly discrepancy analysis for NIDS/OS couples.

\subsection{Semi-automatically generated overlap cases}
\label{sec:semi_automatic_related_works}

Other works focused on chunks overlaps from a different perspective.
Bock et al. \cite{bock2019geneva} used a genetic algorithm named Geneva to find packet sequences differently processed between CS and hosts.
Theoretically, this algorithm can perform evasion attacks with overlapping IPv4 or TCP chunks as it can modify the corresponding header fields and payload.
However, it did not find any such chunk sequences.
Zhang et al.~\cite{zhang2022statediver} derived Geneva to find novel packet sequences that bypass Suricata or Snort.
Their tool, StateDiver, found one (quite complex) successful technique using IPv4 fragmentation and TCP segmentation.
We suppose that Geneva and StateDiver failed to find more overlap-based techniques because 1) successful evasion attempts drove the tools away from this strategy and/or 2) some tested DPIs and hosts may have had the same reassembly policy.
One cannot be sure that all the overlap cases were tested exhaustively.

Di Paolo et al.~\cite{di2023new} also used a fuzzing-like approach to verify OS compliance with the IPv6 specification~\cite{deering2017rfc} when processing overlapping fragments.
They derived the Shankar and Paxson model~\cite{shankar2003active} by permuting and duplicating the chunks, and they found that none of the tested OSes (which were Linux, Windows, or BSD-based) conform.

Wang et al. introduced SymTCP and Themis tools~\cite{wang2020symtcp,wang2021themis}, which both use symbolic execution to find TCP packet
sequences that are processed differently between TCP implementations.
SymTCP successfully found a data overlapping strategy to evade Zeek version 2.6.
However, while this method could theoretically cover all overlap types, SymTCP cannot find exhaustive overlapping test cases because of its incapability to model retroactive behaviors on data buffers (as rightly explained in \cite{wang2020symtcp} section IX).
Themis could not find any TCP-based attack
strategy based on data overlap ambiguity because all the tested implementations were Linux-based.
These implementations may, therefore, have the same TCP reassembly policy. 
In theory, the Themis tool can show discrepancies in reassembly policies if there are any.
However, if none are found, one cannot easily retrieve the reassembly policy.
The authors also highlight that adapting the tool to any OS may require quite important efforts.

\section{Discussions and future works}
\label{sec:discussion}

This section discusses the exploitability of the results described in~\Cref{sec:results}.
It also debates NIDS countermeasures regarding overlap data ambiguity and gives recommendations.

\subsection{Overlap-based attack usability}

\subsubsection{Relation differences}

An attacker that would like to use overlaps to perform an insertion or an evasion attack may struggle differently depending on the relation.
If the goal is an evasion by making the NIDS misassemble the transport header, then the attacker would benefit more from $S$, $Si$, or $Eq$ overlaps because they make the chunks start at the same byte offset.
If these relations are not used, the attacker may need to add a small chunk on the left, which may be considered as "weird" chunks by NIDSes, especially for IP fragments.
Zeek, for instance, considers fragments under 64 bytes as too small, producing a "weird" logging entry.
Differently, suppose the attacker aims to make the NIDS hash calculation fail for a given file. 
Any overlap relation is helpful in that case because one bit flip on the overlapping data portion is enough to change the hash.

\subsubsection{Context importance}
The overlap relation reassembly may change depending on the testing mode, as mentioned in~\Cref{sec:results}.
IP testing exhibits many reassembly differences between the modes.
We hypothesize this is partly due to the increasing importance of the fragment which has the \emph{More Fragments} bit unset with $single$.
If this fragment is dropped, the reassembly conditions are not met, and the test case is ignored.
Differently, the TCP testing mode has much less impact on overlap reassemblies, making the results more adaptable to a larger diversity of segment sequences.
It should not be forgotten, however, that in both testing modes, an extra segment was added before (byte-wise) and sent after (time-wise) the overlap segments.
In future works, we plan to extend overlap cases' \emph{testing context} (e.g., adding an extra non-overlapping fragment after the overlapping chunks).
This should give a more complete picture of OS reassembly policies.

\subsubsection{Chunk sequence alteration before reaching NIDS or supervised host}
\label{sec:chunk_alteration_before}

Offloaded stacks on NIC might impact the OS and NIDS policies described in~\Cref{sec:results}.
NIDS developers advise configuring NIDS instances so nothing alters the supervised traffic, such as NIC offloading.
These recommendations, however, do not guarantee that such alteration does not occur on the supervised hosts themselves.
We thus plan to analyze NIC's impact in future works.

\cite{li2017lib,wang2017your} show that some middleboxes drop or reassemble IPv4 fragments, but what middlebox causes this is unclear (e.g., routers, end host's firewall).
We also plan to test these middlebox reassembly policies to clarify this point.

Finally, due to well-known and unwanted transport issues, chunks may be delayed or dropped, changing the original overlap relation(s) between the chunks.

\subsection{Reassembly policy configurability}

Suricata and Snort allow one to configure reassembly policies according to the supervised host OS, while Zeek does not.
We analyze and compare configurable and non-configurable reassembly policy costs in the following.

\subsubsection{Configurability cost}

Making a NIDS configurable regarding various implementation reassembly policies necessitates several steps.
As OS network protocol implementations may evolve over time, checking whether their policies have changed regularly is necessary.
NIDSes should be able to easily modify and add reassembly policies as well as extend the mapping between OS versions and reassembly policies.
NIDS reassembly policies must be carefully tested to ensure the NIDS reassembles consistently with OSes and that no bug was introduced.
Finally, NIDS users must correctly configure their NIDS instance to associate IP addresses with reassembly policies.
This configuration task is challenging as an organization's IT infrastructure may rapidly evolve and comprise hundreds (or many more) of supervised hosts.
Moreover, IP addresses may be non-static, increasing the human cost of such a configuration even more.
This configuration could be painlessly automated through passive OS fingerprinting~\cite{p0f,holland2021new} or active fingerprinting~\cite{shankar2003active,shamsi2016unsupervised}.
As several OSes may be behind an IP address, NIDSes should consider changing the IP address-based reassembly to a flow-based reassembly.
We plan to investigate these challenges in future works.

\subsubsection{Non-configurability cost}

A NIDS that does not make the reassembly policy configurable must propose another countermeasure to face overlap-based attacks.
For example, an alert-based solution is possible and would consist in raising an alert whenever an inconsistent data overlap is detected (Zeek, Suricata and Snort implement such a countermeasure).
Several approaches may be adopted depending on whether a chunk sequence with overlapping data is inherently considered malicious\footnote{John and Olovsson's work~\cite{john2008detection} analyzed the data consistency of some $Eq$ IPv4 fragment overlaps in 2008. But, to our knowledge, no work has systematically analyzed whether overlapping chunks with inconsistent data are observed in the wild, and if so, inferred the beningness of such chunk sequences. There are, however, benigm reasons for complete or partial overlaps to occur (for example, see~\cite{touch2013rfc,floyd2000rfc2883}).}.
If so, there may not be the need for extra information logging, but if not, such a NIDS must log the beginning and finishing byte offsets as well as data on overlapping portions for further analysis.
In any case, reassembled data from these chunk sequences must not be used (e.g., for TLS certificate extraction) because the NIDS would not know the monitored host reassembled payload.

\subsection{Recommendations for OSes and NIDSes}
\label{sec:reco}

Overlap ambiguity is a long-standing problem, as Ptacek and Newsham~\cite{ptacek1998insertion} initially reported 25 years ago.
OSes have changed their reassembly policies over time.
However, they still exhibit reassembly diversity.
We hypothesize that this diversity is partially caused by the lack of recommendations inside IPv4 and TCP RFCs~\cite{rfc791,rfc9293}.
We recommend that the OSes implement the same policy (e.g., always use original data, ignoring overlapping fragments) so that ambiguities (slowly) disappear with new releases.
Until then, NIDSes with configurable policies must propose multiple reassembly policies and, continuously testing their consistencies.
The NIDSes must especially implement single mode rassemblies because they best describe OS behaviors independently of the testing context.

\section{Ethical considerations}

\subsection{Responsible Disclosure}

We contacted Suricata, Snort, and Zeek developers about NIDS inconsistencies with respect to the latest Windows, Linux, SunOS, and FreeBSD/OpenBSD overlap reassembly policies.
We gave the NIDSes some months to fix the reported issues before submitting the paper.
Snort did not respond to the solicitations, and Zeek acknowledged the results.
The CVE-2024-32867~\cite{cve_ours} was assigned to the Suricata \emph{bsd}-related misassemblies.
We also notified Suricata that we found a display bug during the TCP tests.
In particular, some overlapping chunk payloads appeared twice in the \emph{payload} field of the \emph{eve.json} file (the main logging file).
This, however, does not impact the TCP buffer with which the pattern matching is done.
This bug is now fixed.

\subsection{Censorship Systems}

Improving NIDS security and performance has the side effect of improving censorship systems (CS).
Different techniques may be used to elude CS, such as using a VPN~\cite{nobori2014vpn}, encapsulating or mimicking a non-censored protocol traffic~\cite{meek,obsf4}, inserting a packet that desynchronizes host and censorship-related stateful network traffic analysis tools~\cite{khattak2013towards,bock2019geneva,bock2021even,li2017lib,wang2017your,wang2020symtcp}.
The data overlapping strategy falls into the latest technique.
Some works showed that it was possible to use data overlaps to circumvent CS at least until 2017~\cite{khattak2013towards,wang2017your}, but then, works reported the strategy's unusability~\cite{bock2019geneva,wang2020symtcp}.
Thus, our results should not affect the censorship elusion techniques currently used.
Nonetheless, even if this strategy is in use to circumvent censorship systems, we consider that improving defense capabilities outweigh the negative impacts on censorship elusion techniques.

\section{Conclusion}

In this paper, we adapt well-known \emph{evasion} and \emph{insertion} attack types, refining some specific characteristics related to the overlapping ambiguity.
We propose to use Allen's interval algebra-based modeling to describe chunk sequences and ensure the enumeration exhaustiveness of overlap types.
This enables us to test OS reassemblies completely regarding overlapping pairs of IPv4 and IPv6 fragments as well as TCP segments.
The results show that OS reassembly policies have evolved since the last testing campaigns.
Overall, we demonstrate that 9 (resp. 6) out of 12 IP or TCP reassembly policies are \emph{inconsistent} with the tested OSes for Suricata (resp. Snort). 
Zeek only reassembles consistently with Windows OSes the TCP overlaps.
This exposes these NIDSes to insertion and evasion attacks.
NIDSes with configurable reassembly policies are less subject to attacks, especially segment-based ones, since TCP policies have changed little.
The CVE~\cite{cve_ours} was assigned to the Suricata \emph{bsd}-related misassemblies we uncovered.
Finally, we intend to extend the test context (e.g., multi-chunk overlaps) to completely capture OS reassembly policy complexity as test cases are reassembled differently across testing modes.

\begin{credits}
    \section*{Acknowledgments}
   This work was supported by a grant from the French National Cybersecurity Agency (ANSSI).
\end{credits}

\bibliographystyle{splncs04}
\bibliography{202504_arxiv} 


\end{document}


\begin{tikzpicture}
    \draw[black, thick] (0,0) -- (1.5,0) node [midway, above, sloped, inner sep=1pt] (TextNode) {$X$} ;
    \draw[black, thick] (0,0.5) -- (0.75,0.5) node [midway, above, sloped, inner sep=1pt] (TextNode) {$Y$} ;
\end{tikzpicture}